\documentclass[prd,aps,nofootinbib]{revtex4}
\def\lesssim{{\
\lower-1.2pt\vbox{\hbox{\rlap{$<$}\lower5pt\vbox{\hbox{$\sim$}}}}\ }} 
\def\gtrsim{{\
\lower-1.2pt\vbox{\hbox{\rlap{$>$}\lower5pt\vbox{\hbox{$\sim$}}}}\ }}

\begin{document}

\title{Condition for Primordial Black Hole Formation\\
in Randall-Sundrum Cosmology}
\author{Masahiro Kawasaki}
\affiliation{Research Center for the Early Universe, 
University of Tokyo,
Tokyo 113-0033, Japan}
\date{\today}

\begin{abstract}
We consider  spherical collapse in the
Randall-Sundrum type II model and estimate 
the critical over density for
black hole formation in the radiation dominated era. 
It is found that when (density)$^{2}$-term is 
dominant in the modified Friedmann equation 
the critical density is smaller than in the standard cosmology, 
which implies that PBHs are more easily produced in the 
Randall-Sundrum model.
\end{abstract}

\maketitle

\section{Introduction}

Primordial black holes (PBHs) are e produced 
in the early Universe as a result of initial large density
fluctuations. PBHs are stable against evaporation~\cite{hawking} 
if their masses are small ($\lesssim 10^{15}$~g), 
and hence such PBH is a good candidate for dark matter 
of the Universe.
PBHs with smaller masses evaporate by now and can give significant 
contributions to background $\gamma$-rays and cosmic rays. 
In particular, evaporation 
of PBHs may account for the excess of low energy anti-protons 
observed  by the BESS experiment~\cite{BESS}. 

Since regions with large over density $\delta$ can collapse into
black holes, the number density of PBHs is very sensitive to 
the spectrum of the initial density fluctuations and the critical
over density $\delta_{c}$ above which the black hole is formed. 
In the standard cosmology, the critical density was estimated by
Carr~\cite{carr} for a radiation dominated universe and it was 
applied to a number of PBH scenarios 
(e.g. \cite{carr2,green,KY}, for review see \cite{Carr:2003bj}). 
 
However, the evolution of the early universe may be drastically
changed in the presence of extra-dimensions and branes which are predicted by string theories~\cite{polchinski}. 
In braneworld cosmology, the
standard particles including ourselves are confined in one of
branes and only gravity can travel in a bulk space. 
Randall and Sundrum~\cite{RS} proposed a simple and attractive 
model (called Randall-Sundrum Type II model) for the braneworld,
where a brane with positive tension is embedded in 
a 5D anti-de Sitter space. 
In this model the Friedmann equation is modified and hence 
cosmological evolutions of the scale factor and density are 
different from those in the standard cosmology, which leads to
different evolution of the PBHs after their 
formation~\cite{liddle,majumdar,sendouda}.

In this letter, we consider  spherical collapse 
of a region with large
over density  in the radiation dominated era of the 
Randall-Sundrum model and determine the critical over density 
$\delta_{c}$. It is shown that the critical density is small 
compared with result of the standard cosmology, which suggests 
that more abundance of PBHs is expected in  the Randall-Sundrum 
cosmology.

\section{Spherical Collapse Model in the Standard Cosmology}

First we consider the collapse in the standard cosmology in order
to clarify the difference between the standard and  
the Randall-Sundrum models. 
Let us consider a spherical region with radius $R$ and over density
$\delta$ in a radiation-dominated universe. If the radius is larger
than the horizon, the region behaves like a Friedmann universe. 
Thus, the evolution of $R$ is described by
\begin{equation}
   \label{eq:expansion}
   \left(\frac{dR}{dt}\right)^{2} = 
   \frac{8\pi}{3M^{2}_{4}}\rho R^{2}  - K
\end{equation}
where $t$ is the cosmic time, $M_{4}$ is the 4D Planck mass,
$\rho$ is the radiation density and $K$ is a positive constant.
The solution of this equation is written as 
\begin{eqnarray}
    R & = & A \sin\theta, \\
    t & = & B (1-\cos\theta),
\end{eqnarray}
where $\theta$ is a parameter. From Eq.(\ref{eq:expansion}) the constants $A$ and $B$ is related by
\begin{eqnarray}
    \frac{A^4}{B^2} & = & \frac{8\pi}{3M^{2}_{4}}\rho R^4,\\
    \frac{A^2}{B^2} & = & K.
\end{eqnarray}
Please notice that the radiation density $\rho$ changes as 
$\sim R^{-4}$ and hence $\rho R^{4}$ is constant. 

The spherical region stops expanding when $\theta = \pi/2$, which
gives the critical time $t_c$ and radius $R_c$,
\begin{eqnarray}
    \label{eq:critical-radius}
    R_{c} & = & A, \\
    \label{eq:critical-time}
    t_{c} & = & B.
\end{eqnarray}
For $\theta \ll 1$ $R$ and $t$ are expanded as 
\begin{eqnarray}
    R & \simeq & A(\theta - \theta^3/6 + \ldots ),\\
    t & \simeq & B(\theta^2/2 - \theta^4/24 + \ldots ).
\end{eqnarray}
Then $\theta$ and $R$ can be expressed as a function of $t$ as
\begin{eqnarray}
    \theta^2 & \simeq & \frac{2t}{B} + \frac{\theta^4}{12}
    \simeq \frac{2t}{B} + \frac{t}{6B}\frac{2t}{B},\\
    R & \simeq & A\left(\frac{2t}{B}\right)^{1/2}
    \left( 1+ \frac{t}{12B} -  \frac{t}{3B} \right) 
     \simeq  A\left(\frac{2t}{B}\right)^{1/2}
    \left( 1-\frac{t}{4B}\right).
\end{eqnarray}

On the other hand, the homogeneous background universe expands
as
\begin{equation}
    R_{b} =  A_{b}\left(\frac{2t_{b}}{B_{b}}\right)^{1/2},
\end{equation}
where the subscript ``b'' represents the homoegeneous background 
and 
\begin{equation}
    \frac{A_{b}^4}{B_{b}^2}  =  \frac{8_{b}\pi}{3M^{2}_{4}}
    \rho_{b} R_{b}^4.
\end{equation}
In order to calculate the
over density $\delta = (\rho -\rho_{b})/\rho_{b}$ we must fix the coordinate system or gauge. 
Here we take ``synchronous gauge'', i.e., $t=t_{b}$, $A=A_{b}$
and $B=B_{b}$.  Then, the over density given by
\begin{equation}
    1 + \delta = \frac{R_{b}^4}{R^4} \simeq 1 + \frac{t}{B}.
\end{equation}
Therefore, at the initial time $t_i$ the over-density $\delta_i$ 
is given by
\begin{equation}
    \delta_i = \frac{t_i}{B}.
\end{equation}
Using Eqs.~(\ref{eq:critical-radius}) and (\ref{eq:critical-time}) we
obtain
\begin{eqnarray}
    \frac{t_c}{t_i} & = & \frac{1}{\delta_i},\\
    \frac{R_c}{R_i} & = & \frac{A}{A\theta_i} 
    = \left(\frac{B}{2t_i}\right)^{1/2}
    = \left(\frac{t_c}{2t_i}\right)^{1/2} 
    = \frac{1}{(2\delta_i)^{1/2}}.
\end{eqnarray}

For the region to collapse and become a black hole, the critical
radius must be larger than the Jeans radius $R_{J}$ 
which is given by [see Eq.(\ref{eq:Jeans})] 
\begin{equation}
    R_{J} = \frac{\ell_{H}(t_c)}{\sqrt{6}},
\end{equation}
where $\ell_{H}(t)$ is the particle horizon at $t$ 
and $\alpha (\sim O(1))$
is introduced to take account of ambiguity in defining $R_{J}$. 
By requiring $R_c > R_{J}$ we obtain the condition 
for the black hole formation,
\begin{equation}
    \delta_i > \delta_c \equiv 
    \frac{\alpha^{2}}{3}\left(\frac{2t_i}{R_i}\right)^2.
\end{equation}
Thus, when $R_i=\ell_{H}(t_{i})=2t_{i}$, the critical over-density $\delta_c$ 
is given by~\footnote{
This critical over density is the same as the result in 
Ref.~\cite{carr}. However, this is accidental because 
the different gauge and Jeans radius are used.
If we use ``uniform Hubble gauge'' adopted in Ref.~\cite{carr}
the critical density is given by $\delta_{c}=2\alpha^{2}/3$.}
\begin{equation}
    \label{eq:critical}
    \delta_c = \frac{\alpha^{2}}{3}.
\end{equation}
\section{Spherical Collapse in Randall-Sundrum Model}

Now we study the spherical collapse in Randall-Sundrum model. 
In this model the evolution of the spherical region with radius
$R$ is described by the modified Friedmann equation 
as~\cite{shiromizu}~\footnote{
Here we presume that the spherical collapse is described by
the modified Friedmann equation. 
However, the bulk contribution (5D Wyle tensor)  
might spoil this picture~\cite{shiromizu}. 
Since the evaluation of the bulk contribution is very difficult 
and is beyond the scope of the present letter, 
we simply assume that the bulk contribution is small. }
\begin{equation}
   \label{eq:expansion-RS}
   \left(\frac{dR}{dt}\right)^{2} = 
   \frac{8\pi}{3M^{2}_{4}}\rho R^{2} 
   + \frac{16\pi^{2}}{9M^{6}_{5}}\rho^{2} R^{2} + 
   \frac{\mu}{R^{2}} - K,
\end{equation}
where $M_{5}$ is the 5D Planck mass and $\mu$ represents 
the dark radiation. When the density is 
small ($\rho < 3M_{5}^{6}/(2\pi M_{4}^{2})$) and the dark
radiation is absent, 
Eq.(\ref{eq:expansion-RS}) is the same as the standard one.
However, for the case of large density, $\rho^{2}$-term
dominates the RHS of Eq.(\ref{eq:expansion-RS}) and
the cosmological evolution is quite different 
from that in the standard cosmology.
Therefore, we only consider the latter case. Then, the modified
Friedmann equation is given by
\begin{equation}
   \label{eq:expansion-RS2}
   \left(\frac{dR}{dt}\right)^{2} = 
   \frac{16\pi^{2}}{9M^{6}_{5}}\rho^{2} R^{2} - K,
\end{equation}
In similar way to the standard case, we obtain the solution of
Eq.(\ref{eq:expansion-RS2}) in the radiation dominated era,
\begin{eqnarray}
    R & = & A \sin^{1/3}\theta, \\
    t & = & B \int_{0}^{\theta}d\theta' \sin^{1/3}\theta',
\end{eqnarray}
where the constants $A$ and $B$ is related by
\begin{eqnarray}
    \frac{A^8}{9B^2} & = & \frac{16\pi^{2}}{9M^{6}_{5}}
    \rho^{2} R^8,\\
    \frac{A^2}{9B^2} & = & K.
\end{eqnarray}
The critical time $t_c$ and radius $R_c$ are given by
\begin{eqnarray}
    \label{eq:cri-radius-RS}
    R_{c} & = & A, \\
    \label{eq:cri-time-RS}
    t_{c} & = & -B\frac{\sqrt{\pi}\Gamma(-1/3)}{\Gamma(1/6)}
    \simeq 1.29B \equiv \beta B.
\end{eqnarray}
For small $\theta$, $R(\theta)$ and $t(\theta)$ are expanded as
\begin{eqnarray}
    R & \simeq & A\theta^{1/3}(1 - \theta^2/18 + \ldots ),\\
    t & \simeq & \frac{3}{4}B\theta^{4/3}
    (1 - \theta^2/45 + \ldots ).
\end{eqnarray}
Then, $R$ is expressed in term of $t$ as 
\begin{equation}
  R \simeq A\left(\frac{4t}{3B}\right)^{1/4}
  \left(1 -\frac{1}{20}\left(\frac{4t}{3B}\right)^{3/2}
  + \ldots\right)
\end{equation}
Choosing the same gauge as before, the background universe evolve as
\begin{equation}
  R_{b} = A\left(\frac{4t}{3B}\right)^{1/4},
\end{equation}
which leads to the initial over density
\begin{equation}
  \label{eq:delta-RS}
  \delta_{i} = \frac{1}{5}\left(\frac{4t_{i}}{3B}\right)^{3/2}.
\end{equation}
From Eq.(\ref{eq:delta-RS}) the critical time $t_{i}$ and 
the critical radius $R_{i}$ are 
\begin{eqnarray}
    \frac{t_c}{t_i} & = & 
    \frac{4\beta}{3(5\delta_{i})^{2/3}},\\
    \frac{R_c}{R_i} & = & \frac{1}{(5\delta_{i})^{1/6}} .
\end{eqnarray}
Next, let us consider the condition that the region collapses into
a black hole, $R_{c} > R_{J}$.  The Jeans length can be read 
from the mode equation of the density perturbations $\delta_{k}$
for the modified Friedmann model~\cite{liddle,brax},
\begin{equation}
  \label{eq:mode}
  \ddot{\delta}_{k}
    + H \dot{\delta}_{k}+
    \left[ -\frac{16\pi}{3M_{4}^{2}}\rho 
    -\frac{32\pi^{2}}{M_{5}^{6}}\rho^{2}
    +\frac{1}{3}\left(\frac{k}{a}\right)^{2}\right]\delta_{k} = 0,
\end{equation}
where $a$ is the scale factor and $H$ is the Hubble. From 
Eq.(\ref{eq:mode}) the Jeans length is given by
\begin{equation}
    \label{eq:Jeans}
   \frac{a}{k_{J}} =  \left\{
     \begin{array}{lll}
       \frac{4\sqrt{\pi\rho}}{M_{4}} & 
       = \frac{1}{\sqrt{6}}\ell_{H} &
       ({\rm low}~ \rho )\\[0.8em]
       \frac{4\sqrt{6}\pi\rho}{M_{5}^{3}} & 
       = \frac{1}{\sqrt{6}}\ell_{H} &
       ({\rm high}~\rho )
     \end{array}
     \right.
\end{equation}
Therefore, taking $R_{J} = \alpha/\sqrt{6}\ell_{H}$,
the condition for the black hole formation is written as
\begin{equation}
  \delta_{i} > \frac{8\alpha^{2}\beta^{2}}{135}
  \left(\frac{4t_{i}}{3R_{i}}\right)^{2}
  \simeq 0.10 ~\alpha^{2}\left(\frac{4t_{i}}{3R_{i}}\right)^{2}. .
\end{equation}  
Then, the critical over density at 
$R_{i}=\ell_{H}(t_{i})= 4t_{i}/3$
is given by 
\begin{equation}
   \label{eq:critical-RS}
   \delta_{c} \simeq 0.1~\alpha^{2}.
\end{equation}

\section{Conclusion}

From Eqs.(\ref{eq:critical-RS}) and (\ref{eq:critical}), 
the critical value in Randall-Sundrum model is about 3 times
smaller than in the standard cosmology.
Thus, PBHs are more easily produced when 
the $\rho^{2}$-term controls the cosmic expansion in the 
Randall-Sundrum model. This can be explained by the growth rate
of the density fluctuations. 
As is seen in Eq.(\ref{eq:delta-RS}) the density perturbations 
grows as $\sim t^{3/2} $ which is faster than $\sim t$ for
the standard case.  

When the distribution for the density fluctuations
is Gaussian, the mass fraction $f(M)=\rho_{\rm BH}(M)/\rho$ of PBHs
with mass $M$ is given by~\cite{green}
\begin{equation}
  f(M) = \int_{\delta_{c}}\frac{d\delta}{\sqrt{2\pi}\sigma(M)}
  \exp\left(-\frac{\delta^{2}}{2\sigma(M)^{2}}\right)
  \simeq \sigma(M)
  \exp\left(-\frac{\delta_{c}^{2}}{2\sigma(M)^{2}}\right),
\end{equation}
where $\sigma(M)$ is the mass variance at horizon crossing and
we assume $\sigma(M) \ll \delta_{c}$. This shows that the abundance
of PBHs is very sensitive to $\delta_{c}$. Therefore, the smaller 
critical value (\ref{eq:critical-RS}) gives a significant effect on 
the mass spectrum of PBHs.
Suppose that the $\rho^{2}$-dominated era ends at
$t=t_{*}$. The horizon mass $M_{*}$ at $t_{*}$ is $\sim 
\rho_{*}t_{*}^{3}$. Since the mass of the produced PBH is $\sim$ 
horizon mass, PBHs with mass smaller than $M_{*}$ are produced
in $\rho^{2}$-dominated era and have much larger abundance 
than more massive PBHs. This fact may be useful 
when one searches  the  extra-dimension by observing 
$\gamma$-ray and anti-protons from evaporation of PBHs.

Finally we make a comment on the choice of $R_{i}$ in 
estimating the critical density, In this letter we take
$R_{i}=\ell_{H}(t_{i})$. Another reasonable choice is  
$R_{i}=H^{-1}(t_{i})$, In this case, the difference between
$\delta_{c}$(standard) and $\delta_{c}$(Randall-Sundrum) becomes
larger.  So the conclusion that the PBHs formation is easier 
in Randall-Sundrum cosmology does not change.

In summary, we have considered the spherical collapse in 
Randall-Sundrum model and estimated the critical over density for
the black hole formation in the radiation dominated era. 
It has been found that when the Hubble expansion is dominated by
$\rho^{2}$-term in the modified Friedmann equation the critical
density is smaller than in the standard cosmology, which implies 
that PBHs are more easily produced in the Randall-Sundrum model.

\end{document}